\documentstyle[preprint,aps,prd]{revtex}

\begin{document}

\title{
Phenomenological theory of mode collapse-revival in  confined Bose gas}
 
\author{L.P. Pitaevskii}
 \address{Department of Physics, Technion, 32000 Haifa, Israel\\ 
and\\ 
Kapitza Institute for Physical Problems, 117454 Moscow, Russia}
 
 \date{January 2, 1997}
 
 \maketitle
 
 \begin{abstract}
A phenomenological theory of mode collapse-revival 
for a system with a weak nonlinearity is presented. The theory takes into 
account  
fluctuations of the number $n$ of quanta of oscillations. The collapse
time $\tau _c ^{-1}$ for a mode of frequency $\omega _0$ turns out to be  
$\omega _0(\sqrt{n/2}\hbar \omega _0 \mid \nu \mid)$, 
where $\nu$ is a coefficient in a nonlinear correction to 
the mode frequency with respect to the oscillation energy $E$.  For a Bose 
gas in a harmonic trap $\tau _c $ is order of 
$\omega _0^{-1}\sqrt{\mu N/E\hbar \omega _0}$.
This value is order of 250 ms. for typical experimental conditions.
The coefficient $\nu$ is calculated for the breathing mode in an isotropic 
trap.
 \end{abstract}
 

\narrowtext
The recent discovery of Bose-Einstein condensation of alkali atoms  
 \cite{1}
confined in magnetic traps has opened a new important field 
 of investigation of quantum 
phenomena on a macroscopic scale. The most interesting  
possibility is to observe specific quantum phenomena which have no classical 
analogy. One  such  phenomenon is the collapse and revival of  
coherent quantum states recently observed for an atom in a 
electromagnetic field in a resonant cavity \cite{2} . The problem has a 
developed theoretical background, and for a comprehensive discussion see in 
Ref. \cite{the}. 

Several significant  works have been carried out recently about 
manifestation 
of this phenomenon in Bose-condensed gases. In paper \cite{3} 
properties of a ground state are discussed. Kuklov et. al. \cite{bir} 
considered dephasing  of a finite amplitude oscillation in a 
finite Bose system because of atom occupation number fluctuations in the 
oscillation. They suggested that this mechanism can explain the observed
damping of oscillations and
used a microscopic model to calculate the collapse time and they concluded 
that the theory 
can give a reasonable order of magnitude of this time compared to 
the damping observed in experiments \cite{cor},\cite{ket}. In \cite{4} 
the authors considered the dephasing  because of the zero-point 
fluctuations of occupation numbers of atoms. Since the authors have 
withdrawn 
the paper we don't discuss it in detail. Note only that we believe 
that zero-point fluctuations cannot result in the effect under consideration.

In this paper we will present a general phenomenological theory of 
collapse-revival of a highly 
excited mode of oscillation of  a superfluid Bose system. We will take 
into account not fluctuations of occupation numbers of atoms but rather 
fluctuations in the numbers  of quanta of oscillations of the mode. (This is 
by no means 
the same for  low-frequency oscillations.) We believe that this choice 
of variables is more proper then used in \cite{bir}. The physical meaning 
of the effect is that a coherent oscillation is a linear combination of 
 stationary states of a system and the number of quanta is a natural 
specification of a stationary state of an oscillator. In this way one 
doesn't meet the troublesome 
problem of  diagonalization of the Hamiltonian. Of 
course a problem can be solved in any variables and a 
microscopic theory always gives more information about  some mechanism of 
the effect.  
From another side, the phenomenological approach is more transparent and 
gives direct possibility for comparison with an experiment.  

According to general theory \cite{the} the effect in consideration 
is based on 
the fact that transition frequencies for stationary component of a 
coherent state of the system are different. This effect doesn't exist for 
a purely harmonic oscillator and hence is completely  defined by  
non-linearity. 
In \cite{the} a two-level system in an electromagnetic field has been 
considered. One may say that this is a
limiting case of an extremely  strong nonlinearity. For our system however 
the opposite case of weak nonlinearity is more proper. 

First of all we will present the collapse-revival theory in a 
form more suitable for our purpose.
In the case  of weak nonlinearity the frequency 
of the oscillation mode under consideration can be written as:

\begin{eqnarray}
\omega = \omega _0+\delta \omega = \omega _0 (1+\nu E) ,
 \label{1}
 \end{eqnarray}
where $E$ is an energy of oscillations, and 
$\mid \nu \mid E <<1$ (weak non-linearity). Stress that constant $\nu $ 
describing the first nonlinear correction has a  purely classical meaning. 
For an anharmonic oscillator with a Hamiltonian 
$H=p^2/2m + m\omega _0^2x^2/2+\alpha x^3/3+\beta x^4/4$ one has:
\begin{eqnarray}
 \nu =-\frac{5\alpha ^2}{6m^3 \omega _0 ^6}+\frac{3\beta }{4m^2\omega _0 
^4}. 
\label{osc}
\end{eqnarray}
Taking into account that $\hbar \omega = (\partial E_n/\partial n)$, we 
can rewrite
this equation  
in the quantum form 
\begin{eqnarray}
\omega _n = E_n/\hbar=\omega _0n +bn^2/2 ,
 \label{2}
 \end{eqnarray}
where $n$ is the number of quanta in the given excited state of the system 
($n>>1$), $E_n$ is the energy of the state and
\begin{eqnarray}
b=\hbar \omega _0^2 \nu.
 \label{3}
 \end{eqnarray}
 Let us consider now a coherent state of our oscillator. Its wave 
function can be presented in the well-known form:
\begin{eqnarray}
\psi = C\sum _n c _n \psi _n exp (-i\omega _n) ,\mid c _n \mid ^2 = 
exp(-\bar{n}) \bar{n}^n/n!,
 \label{4}
 \end{eqnarray} 
where $\psi _n $ is a wave function of a stationary state of the 
oscillator where $n$ quanta are exited, $\hbar \omega _0 \bar{n} = E$. All 
quantities besides $\omega _n $ can be taken for a harmonic oscillator.
In our case:
\begin{eqnarray}
\bar{n} >>1, \mid c _n \mid ^2\approx \frac{1}{\sqrt{2\pi 
\bar{n}}}exp \left (-\frac{(n-\bar{n})^2}{2\bar{n}}\right ).
 \label{5}
 \end{eqnarray}
Let us calculate now an average value of the oscillator coordinate $x$ 
using the wave function (\ref{4}). Taking into account that only 
transitions $n \rightarrow n \pm 1$ are important, one gets:
\begin{eqnarray}
<x(t)> \approx \sum _n\mid Cc _n \mid ^2 cos((\omega _0 +bn)t) . 
 \label{6}
 \end{eqnarray}
For small enough values of $t$ we can replace the summation over $n$ by 
integration. The result is Gaussian damping of amplitude of the 
oscillation according to:
\begin{eqnarray}
<x> \sim exp(-\bar{n} b^2t^2/2) \equiv exp(-(t/\tau _c)^2),  
 \label{7}
 \end{eqnarray}
where
\begin{eqnarray}
\tau _c ^{-1}= (\sqrt{\bar{n}/2}\mid b \mid )=\omega _0 
(\sqrt{\bar{n}/2}\hbar 
\omega _0\mid \nu \mid)=\omega _0 (\sqrt{E \hbar \omega/2}\mid \nu \mid )
 \label{8}
 \end{eqnarray}
The periodicity of expression (\ref{6}) gives immediately the 
revival period:
\begin{eqnarray}
\tau _r =\frac{\pi}{\hbar \omega _0^2\mid  \nu \mid }
 \label{9}
 \end{eqnarray}
Note first of all that amplitude of the oscillations is proportional to 
$\sqrt{ \bar{n}} $. Thus our phemenological theory gives the same amplitude 
dependence of collapse time as the theory of Ref. \cite{bir}.
We however didn't use so far any information about microscopic properties of 
our system. 
It means that the effect is not related in our description to the fact 
of Bose condensation. 
Every system possessing excitations with long enough lifetime is 
suitable. The coefficient $\nu $ can be calculated or measured and thus the 
theory contains no fitting parameters.

 According to Eq.(\ref{8}) the collapse time $\tau _c$ decreases 
with 
increase of $\bar{n} $. One must however take into account still that 
there is a restriction on the oscillation energy $\mid \nu \mid E <<1$, or:
\begin{eqnarray}
1<<n<<\frac{1}{\hbar \omega _0\mid \nu \mid }
 \label{n}
 \end{eqnarray}

 This inequality 
ensures applicability of the approximation of weak nonlinearity. It gives :
\begin{eqnarray}
\tau _c >> \tau _{min} \equiv \frac{1}{\omega _0 \sqrt{\hbar \omega _0 
\mid \nu \mid }}.
 \label{10}
 \end{eqnarray} 
The effect can be observed if this $\tau _{min} $ is less that the 
oscillation damping due to dissipation 
which is always present.
The theory we presented is a rigorous one at conditions (\ref{n}).
To apply the theory to our problem of oscillations of
Bose condensed gas in a harmonic magnetic trap
one must calculate the nonlinearity  coefficient 
$\nu $.  
According to Stringari 
\cite{str}, the low frequency 
oscillations with frequency $\omega _0 \sim \omega _H$ are 
well described by equations of hydrodynamics if the number of atoms $N$ is 
large enough. ($\omega _H$ is a frequency of oscillation of an isolated 
atom in the trap.) To observe the effect under consideration here one must 
try to 
have as small a number of atoms as possible. We however will use the 
hydrodynamics approximation taking into account that it works well for the 
present experimental conditions.

Calculation of the frequency shift (\ref{1}) is straightforward but 
the procedure is
a little bit cumbersome, because this shift appears at 
equations for third order terms with respect to the oscillation amplitudes.
(Some results about amplitude dependence of the mode frequencies have been 
presented in Ref \cite{non}.) 
Having in view to estimate the order of magnitude of the effect we will 
restrict ourselves to the consideration of a spherical-symmetric breathing 
mode in an isotropic trap only. In this case the hydrodynamic velocity 
$\bbox{v}$ has the form $\bbox{v}(\bbox{r})=u(t)\bbox{r}$. Introducing a new 
variable $w$ according to $u=\dot{w}/w $ one can reduce the equation 
for $w$ to the form \cite{non}, \cite{cas}: 
\begin{equation}
\ddot{w}  + \omega_{H}^2 w - \omega_{H}^2 /w^4 
= 0 \; .
\label{w}
\end{equation}
Substitution $w(t)=1+x(t)$ and expansion with respect to $x$ gives:
\begin{equation}
\ddot{x}  + 5\omega _H^2 x= 20\omega _H^2x^2-120\omega _H^2x^3.
\label{x}
\end{equation}
One has from (\ref{x}):
\begin{equation}
x \approx Acos(\omega _0t), \omega _0=\sqrt{5}\omega _H, \delta 
\omega/\omega _0 = (7/6)A^2. 
\label{do}
\end{equation}
(In fact $A$ is a fractional amplitude of the oscillations in the cloud 
radius.)

The energy of oscillation can be calculated as twice  the mean kinetic 
energy,
$E=\int \rho _0 \bar{\bbox{v}^2}dV$, where $\rho _0(r)$ is the equilibrium 
density of condensate in a trap. Calculation gives $E= (3/7)N\mu A^2$, 
where $\mu$ is chemical potential of the gas, calculated in the 
Thomas-Fermi approximation. Thus the nonlinear shift coefficient $\nu $
turns out to be:
\begin{equation}
\nu =\frac{49}{18}\frac{1}{\mu N}.
\label{nu}
\end{equation}
This gives finally  the collapse time for the mode under consideration:
\begin{equation}
\tau _c  = \frac{18\sqrt{2}}{49}\omega _0 ^{-1}\frac{\mu 
N}{\sqrt{E\hbar \omega 
_0}}=\omega _0^{-1}\frac{0.79}{A}\sqrt{\frac{\mu N}{\hbar \omega _0}} 
\label{fin}
\end{equation}
It is reasonable to think that equation (\ref{fin}) gives the correct 
order of magnitude
for different collective modes, and holds even for a small number of atoms 
where the hydrodynamic approximation cannot be justified rigorously.

Let us estimate $\tau _c$ assuming typical experimental conditions of the 
paper 
\cite{cor}: $N=4500$, $\omega _0/2\pi = 264$ Hz, 
$A=0.2$. 
According to calculations of Ref. \cite{dal} $\sqrt{\mu N/\hbar \omega 
_0}=104$. 
Then
$\tau_c=246 $ ms.
The authors of Ref. \cite{cor} reported that the lifetime of the low $m=0$ 
mode 
is 110 ms. The calculated collapse time is in reasonable order of 
magnitude agreement with the observed   value, however
both experiments \cite{cor} and calculations \cite{non} 
demonstrate absence of the amplitude dependence of frequency for this mode. 
Thus damping of this mode cannot be explained by the present version of 
the theory. 
Experimental conditions of 
\cite{ket} are less favourable for observation of the collapse because of 
the larger number of the condensed atoms. 

One  must take into 
account also that there is a different reason for nonlinear effects 
in our system because of trivial nonharmonic corrections to the magnetic 
trap confining potential which must be considered too.

More accurate calculations are in progress now.
I hope that these calculations will disclose a mode which will be most 
suitable for observation of this effect.
 I would
like to stress here that the first step to confirm the
presented scenario of the damping is to discover predicted amplitude
dependence. 

It is reasonable to look for other suitable systems to apply 
the theory developed here. Liquid helium clusters and small  piezoelectric 
samples are possible candidates.

	In conclusion a general phenomenological theory of mode collapse and
revival for a system with weak nonlinearity is presented. 
The theory doesn't contain any fitting parameters.
The collapse 
time is expressed through a classical   nonlinearity coefficient 
of the system.
This coefficient  is calculated for the breathing mode of the Bose 
condensate in 
a isotropic trap. The collapse time turns out to be about 250 ms for 
typical experimental conditions.

I would like to thank J. L. Birman for various enlightening discussions.

\end{document}